%% file: main.tex
\documentclass[10pt,conference]{IEEEtran}
\IEEEoverridecommandlockouts

\usepackage{cite}
\usepackage{amsmath,amssymb,amsfonts}
\usepackage{algorithmic}
\usepackage{graphicx}
\usepackage{textcomp}
\usepackage[table]{xcolor}
\usepackage{todonotes}
\usepackage{longtable}
\usepackage{booktabs}
\usepackage{array}
\usepackage{tabularx}
\usepackage{xspace}
\usepackage{tikz}
\usepackage{stfloats}

\usetikzlibrary{positioning,arrows.meta,calc}

\newif\ifdraft
\draftfalse %

    \def\BibTeX{{\rm B\kern-.05em{\sc i\kern-.025em b}\kern-.08em
        T\kern-.1667em\lower.7ex\hbox{E}\kern-.125emX}
    \fi
}

\definecolor{kwP}{HTML}{9D8FE6}
\definecolor{kwAI}{HTML}{62C2C4}
\definecolor{kwCode}{HTML}{EDB94F}
\definecolor{kwApp}{HTML}{DE6F80}
\newcommand{\kwicon}[2]{{#2\,\raisebox{-0.3ex}{\includegraphics[height=1.85ex]{figures/icons/#1.pdf}}}}
\newcommand{\WP}{\kwicon{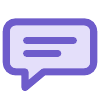}{writing prompts}\xspace}
\newcommand{\RAI}{\kwicon{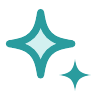}{reading AI output}\xspace}
\newcommand{\RC}{\kwicon{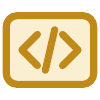}{reading code}\xspace}
\newcommand{\VA}{\kwicon{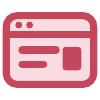}{viewing the app}\xspace}

\newcommand{\aw}{AW\xspace}

\begin{document}

\title{The Vibe Paradox: What EEG Reveals About \\Your Brain When Vibe Coding\vspace{-0.3\baselineskip}}

\author{\IEEEauthorblockN{Olivia Schutz\textsuperscript{1*},
Rudrajit Choudhuri\textsuperscript{1*},
Yunhan Qiao\textsuperscript{1},
Fatima A. Moussaoui\textsuperscript{1},
Julia Russo\textsuperscript{1},\\
Zoe Steine-Hanson\textsuperscript{2},
Anita Sarma\textsuperscript{1}}
\IEEEauthorblockA{\textsuperscript{1}Oregon State University, Corvallis, OR, USA\\\{schutzo, choudhru, qiaoy, moussaof, russoju, anita.sarma\}@oregonstate.edu\\
\textsuperscript{2}OpenBCI, Inc., Brooklyn, NY, USA, zoe@openbci.com\\
\textsuperscript{*}Co-first authors}}

\IEEEaftertitletext{\vspace{-1.1\baselineskip}} %
\maketitle

\begin{abstract}
\textit{Background}: Much of the enthusiasm for vibe coding rests on two intuitive beliefs: (1) that developers stay invested even when they no longer write the code, and (2) that delegating more of the work to AI lowers their cognitive load.

\textit{Objective}: Our goal was to investigate whether developers’ neurophysiological states and interactions with AI are consistent with these beliefs when they work with agentic vs. non-agentic AI on their own project vs. a legacy codebase.

\textit{Method}: We conducted an exploratory within-subjects crossover study with 11 participants across 22 sessions. Each participant used GitHub Copilot Chat in one session and Cursor Agents in the other, working on their own project and a legacy codebase, with tool--codebase pairings randomly assigned and session order counterbalanced. We recorded participants’ EEG, on-screen activity, and prompts, and analyzed 7,924 EEG data points, 984 activities, and 141 prompts.

\textit{Results}: Our findings contradict both beliefs. With the agent, participants remained indifferent to task outcomes, even on their own projects, and cognitive load was higher than with non-agentic AI. Furthermore, their ways of interacting shifted: with non-agentic AI, participants asked how the code worked and read it to verify changes; with the agent, they described desired app behavior and checked results in the running app. 

\textit{Conclusion}: Our study reveals a \emph{vibe paradox}: delegating to an agent left participants motivationally disinterested, even on their own projects, without relieving cognitive load, while coinciding with increased trust in AI. The community should treat this paradox as a first-class concern if developers and their AI agents are to go far together, and not just fast.

\end{abstract}


\input{sections/01-introduction}

\input{sections/02-related-work}

\input{sections/03-methods}

\input{sections/06-threat}

\input{sections/04-results}

\input{sections/05-discussion}

\section*{Acknowledgment}
We thank OpenBCI for facilitating this project and the participants for their time. We also thank Annie Vella for an insightful conversation that helped shape the discussion of our findings.

\bibliographystyle{IEEEtran}
\bibliography{bib}

\end{document}

%% file: sections/01-introduction.tex
\section{Introduction}\label{sec:intro}

Consider two development tasks.
In the first, you add a feature to a project you built yourself, and a coding agent plans the change, edits the files, and runs the code while you describe what you want.
In the second, you fix a bug in a legacy codebase you have never seen, and an AI answers your questions while you make every edit yourself.
\textit{In which task would you feel more engaged? Which would take more effort?}

The first task reflects the promise of vibe coding: developers describe what they want and let AI handle the implementation.
Karpathy, who coined the term, described developers reaching a point where they can ``forget that the code even exists''~\cite{karpathy2025vibe}.
In fact, Pimenova et al.\ found that developers reported reduced cognitive effort when they vibe coded~\cite{pimenova2025good}.
Indeed, much of the enthusiasm for vibe coding rests on two intuitive beliefs: (1) that developers stay invested in their own projects even when they no longer write the code, and (2) that delegating more of the work to AI lowers their cognitive load.

However, neither belief has been tested against what happens with developers as they work; developers' own reports are often a poor guide.
For instance, in one study, experienced developers took 19\% longer with AI, while believing it had made them 20\% faster~\cite{becker2025measuring}.
Thus, testing either belief requires objective measures that do not depend on what developers believe about their own work.

Electroencephalography (EEG) offers such measures, continuously and without interrupting work~\cite{fritz2014using}.
Kosmyna et al.\ used EEG to investigate what happens in users' brains when they write essays with an LLM, and reported weaker brain connectivity than among those writing with a search engine or unaided, along with lower ownership of their essays~\cite{kosmyna2025your}.

Vibe coding makes this inquiry both harder and more pressing~\cite{sarkar2025vibe,sapkota2025vibe}.
A change to code must cohere with the rest of the codebase, including dependencies the developer may not be familiar with.
Moreover, working with an agent is iterative, i.e., developers direct changes, verify them, revert some, and revise their requests over many turns, a different cadence of work from writing an essay.
As a result, their effort and their investment in the outcome may fluctuate as the work unfolds.
This inquiry is especially important now, because vibe coding has lowered the barrier to producing software~\cite{sarkar2025vibe}, so that end-user developers, who write software for their own purposes~\cite{ko2011state}, can now produce it as well.
While EEG has been used to study programming tasks such as code comprehension~\cite{medeiros2021can}, to the best of our knowledge, no work has studied people's neurophysiological states while they vibe code and what producing software this way costs to them. 

In this paper, we thus ask:

\begin{list}{}{\settowidth{\labelwidth}{\textbf{RQ1:}}\setlength{\labelsep}{0.5em}\setlength{\leftmargin}{\dimexpr\labelwidth+\labelsep\relax}\setlength{\topsep}{4pt}\setlength{\itemsep}{3pt}\setlength{\parsep}{0pt}}
    \item[\textbf{RQ1:}] \textit{How do participants' neurophysiological states differ when they work with agentic versus non-agentic AI, on their personal project versus a legacy codebase?}
    \item[\textbf{RQ2:}] \textit{How do participants' interactions differ across these conditions?}
\end{list}

To investigate these questions, we conducted an exploratory within-subjects crossover study in which 11 participants vibe coded with non-agentic AI (GitHub Copilot Chat~\cite{github2026copilot}) and an agent (Cursor Agent~\cite{anysphere2026cursor}).
Each participant used one AI type on their own project and the other on a legacy codebase, with the pairing randomized and session order counterbalanced.
We recorded 7,924 EEG instances, 984 on-screen activity transitions, and 141 prompts across 22 sessions.

Our findings suggested that with the agent, participants' approach--withdrawal motivation (\aw), an EEG measure of a person's tendency to be motivationally invested in the current activity~\cite{coan2004frontal,harmonjones2018role}, stayed near neutral, even on their own projects, and that the agent brought no relief in cognitive load~\cite{gevins2003neurophysiological}, which was, if anything, higher than with non-agentic AI (RQ1).
Furthermore, participants' ways of interacting changed with the AI type.
With non-agentic AI, they worked at the level of the code, asking how it worked and reading it to check changes.
With the agent, they moved up a level of abstraction, where their interactions centered on the app's behavior and outcomes (RQ2).
Interestingly, participants who used the agent on the legacy codebase (and non-agentic AI on their own projects) also reported higher trust in AI than those with the reverse pairing.

Most critically, these effects together surface what we call the \emph{vibe paradox}: delegating to an agent (1) caused participants to be indifferent to outcomes, even on their own projects, (2) brought no relief in cognitive load, and (3) coincided with higher trust in AI.
We frame this motivational indifference as \emph{interest atrophy}, a weakening of investment in outcomes that participants no longer produced themselves.
We argue that what agents save developers in keystrokes, they may charge to developers' motivational investment and cognitive load.

Our work makes three contributions: (1) evidence of developers' neurophysiological states when they vibe code with agentic and non-agentic AI on different projects; (2) an account of how participants' interactions and ways of working changed with the agent, surfacing signs of material disengagement~\cite{sarkar2025vibe} in software development; and (3) the vibe paradox, interpreted through supervisory control~\cite{sheridan1992telerobotics}, reward from one's own choices~\cite{yeung2005erp,haggard2017sense}, motivational intensity~\cite{gendolla2010effort}, and automation complacency~\cite{parasuraman2010complacency}.

%% file: sections/02-related-work.tex
\section{Related Work}\label{sec:related}

Studies of how developers work with AI have mostly examined code completion and chat.
Barke et al.\ found that developers used Copilot in an acceleration mode, to reach a solution they already had in mind faster, and in an exploration mode, to explore options when they were unsure how to proceed~\cite{barke2023grounded}.
Mozannar et al.\ found that thinking about and verifying Copilot's suggestions took the largest share of developers' time~\cite{mozannar2024reading}.
Agents shift this work.
Developers needed less effort to finish tasks with an agent than with a copilot, but did not always understand what the agent had done~\cite{chen2025code}, and in real-world chat sessions they delegated diagnosis, comprehension, and validation to the AI~\cite{tang2026programming}.
Sarkar and Drosos describe vibe coding as material disengagement, in which the developer stops working the code and instead orchestrates its production through the AI~\cite{sarkar2025vibe}.
Vibe coders who relied more on AI re-prompted with the same intent until something worked~\cite{chou2026building}, whereas experienced developers kept control of design and implementation~\cite{huang2025professional}.
Like these studies, we code developers' activities and prompts, but we compare where the same developers check the AI's work, in the code or in the running app, with each AI type and on code that is and is not their own, and we relate these interactions to EEG.

The enthusiasm for vibe coding rests largely on what developers say about it.
Karpathy, who coined the term, described accepting the AI's changes without reading the diffs~\cite{karpathy2025vibe}, and developers report that vibe coding brings flow and joy and reduces cognitive effort~\cite{pimenova2025good}.
Such reports can diverge from measurement.
In a randomized trial, experienced open-source developers believed AI had made them 20\% faster when it had made them 19\% slower~\cite{becker2025measuring}.
We examine whether the two beliefs behind vibe coding, that developers stay invested and that delegating lowers their load, hold when developers' states are measured during the work.

Other studies have measured what delegating to AI costs.
Projects that adopted Cursor gained development velocity briefly but accumulated lasting static-analysis warnings and code complexity~\cite{he2026speed}.
Students who built a website with an agent understood their code less and were less able to extend it than students who used a chatbot, and low-effort interactions, such as auto-accepting edits, were linked to lower comprehension~\cite{balepur2026paired}.
Developers with an AI assistant wrote less secure code while believing it to be more secure~\cite{perry2023users}.
Because these studies rely on post-task code metrics, comprehension tests, and self-reports, they show that delegation has costs but not how invested developers were, how much load they carried, or how they checked the AI's changes while they worked, which our EEG and interaction coding capture.

Physiological measures can capture developers' states as they work.
In software engineering, they have been used to assess task difficulty~\cite{fritz2014using}, emotions and progress~\cite{muller2015stuck}, flow experience~\cite{knierim2025exploring}, and cognitive load during code comprehension~\cite{medeiros2021can}, all without AI, and eye tracking has shown how developers validate and repair Copilot-generated code~\cite{tang2024study}.
Outside programming, people who wrote essays with an LLM showed weaker brain connectivity on EEG than those who used a search engine or no tool, and they reported less ownership of their essays~\cite{kosmyna2025your}.
We follow this work in measuring developers' states during the task, using EEG measures of approach--withdrawal and cognitive load~\cite{coan2004frontal,gevins2003neurophysiological}.
To our knowledge, no study has measured developers' neurophysiological states while they vibe code with an agent, or compared these states on their own code and on code new to them.

%% file: sections/03-methods.tex
\section{Method}\label{sec:method}

Our goal was to describe how participants' neurophysiological states (RQ1) and their interactions with AI (RQ2) differ when they work with agentic versus non-agentic AI, on their personal project versus a legacy codebase.
To do so, we conducted an exploratory within-subjects crossover study in which $n=11$ participants completed 22 sessions (see Fig.~\ref{fig:workflow}).
To answer RQ1, we recorded and analyzed participants' neurophysiological states with EEG; to answer RQ2, we analyzed their interactions with the AI through screen recordings and prompts (Sections~\ref{sec:data} and~\ref{sec:analysis}).
Our institution's IRB approved the study.

\input{figures/workflow}

\subsection{Study Design}\label{sec:design}

Because we studied vibe coding~\cite{karpathy2025vibe,sarkar2025vibe}, participants worked with AI in every session.
To examine the two beliefs in Section~\ref{sec:intro}, we varied who made the changes to the code, the participant or the AI (\emph{AI type}), and whether participants worked on their own project (\emph{project type}).

With \emph{non-agentic AI}, participants used GitHub Copilot Chat~\cite{github2026copilot}, which answered questions and proposed changes that participants could apply themselves.
Inline code completion was also enabled.
With \emph{agentic AI}, participants used the Cursor agent~\cite{anysphere2026cursor}, which inspected the project and edited files directly.
Both tools were model-agnostic interfaces, in which the user could select the language model behind the tool, and participants could choose any model that each tool offered.
AI type therefore varied how much autonomy the AI had in making changes to the code, independently of any particular model.
Each session started from a fresh instance of the tool with its memory of earlier sessions cleared, so every participant began from the same state.
For project type, each participant worked once on their own project and once on a legacy codebase (Section~\ref{sec:tasks}).

Each participant completed two 20-minute sessions, a length recommended by OpenBCI for the headset, in a crossover design in which each participant used both AI types and worked on both project types (Fig.~\ref{fig:workflow}).
Participants were randomly assigned to one of two groups: Group A ($n=5$) used non-agentic AI on the legacy codebase and agentic AI on their own project; Group B ($n=6$) used the reverse pairing.
We counterbalanced session order within each group.

\subsection{Participants}\label{sec:participants}

We recruited participants between May and August 2026 by inviting 1,260 students through an undergraduate software engineering course, the university's recruitment list for online-program students, and snowball sampling.
Participants were required to be at least 18 years old, to have programming and SE experience alongside experience with AI tools, and to provide a personal software project, in any programming language, that could be built and executed, with one or more planned changes they could attempt within 20 minutes.
Because the headset's dry electrodes required direct scalp contact, participants also had to be able to wear the headset with an appropriate fit, work without glasses, and remove earrings or head coverings.
Twenty-two of those invited completed the screening survey.
We reviewed each respondent's project and planned changes against these criteria before inviting them to the study, and 16 enrolled and attended a session.
Four could not complete the study: three could not achieve a stable headset fit (hair texture for two, eyewear for one), and a lab computer failure ended one session.
Twelve participants completed both sessions (see participant demographics in the supplemental material~\cite{supplemental}).
We excluded one of them because of a substantial time gap in their EEG data.
Participants who completed the study received a \$60 Amazon gift card.

\subsection{Tasks}\label{sec:tasks}

\textbf{Personal project.}
Each participant provided a software project they had developed themselves.
Before the session, they listed features they wanted to implement with observable completion criteria.
We set up each project to build and execute on the laboratory computer before the session.
To measure task performance, we checked each listed feature against its observable completion criteria and computed the proportion of listed features the participant completed.

\textbf{Legacy codebase.}
The legacy codebase had to be new to the participants and the models, so we used a non-public web application for speedgolf (name withheld for review).
Because its source is not public, it is unlikely to have appeared in the models' training data.
Participants completed one feature in it: a form for entering a new tournament's details, such as its name, host, and dates, and a button that saves the tournament and exits the form.
Instead of written requirements, participants received screenshots of how the application should look once the feature worked.
To measure task performance, we checked each participant's work against a set of 22 automated test cases and counted the test cases it passed; a test case with two parts received half credit when one part passed.

\subsection{Apparatus}\label{sec:apparatus}

\textbf{Computer and software.}
Participants worked on a Windows 11 laboratory computer.
For non-agentic AI, Copilot Chat ran inside Visual Studio Code; for agentic AI, participants used the Cursor agent on the Pro plan.
By default, Copilot Chat selected its model automatically and Cursor used Claude Opus 4.6, unless participants chose another model.
Participants were not allowed to use any generative AI other than the tools provided.

\textbf{EEG headset.}
We recorded EEG with the OpenBCI Galea Beta~\cite{galea_eeg_sensors}, a head-mounted device that combines EEG with electrooculography (EOG) and motion sensing by an inertial measurement unit (IMU) (see Fig.~\ref{fig:headset}).
Its 10 active dry electrodes need no conductive gel, keeping setup short, and sit at positions of the international 10--20 system: a frontal pair (F1, F2), three central sites (C3, Cz, C4), three parietal sites (P3, Pz, P4), and two occipital sites (O1, O2).
It samples EEG and EOG at 250\,Hz and motion at 50\,Hz.
We used the EOG channels to remove eye-movement artifacts and the motion sensors to reject windows with head movement.
We synchronized the EEG and screen recordings by aligning the task-start timestamps.

\input{figures/eeg}

\subsection{Procedure}\label{sec:procedure}

We collected data between May 22 and September 2, 2026.
Before data collection, we conducted five sandbox sessions to test the equipment and procedure, and we began data collection once a sandbox session required no further changes.
We excluded the sandbox data from all analyses.
After providing informed consent, participants followed the same procedure for each task (Fig.~\ref{fig:workflow}).
Before any tasks, participants reported their background and experience with AI. 
Before each task, a researcher reviewed the requirements with the participant and asked them to restate the requirements in their own words to confirm their understanding.
We fitted the EEG headset and checked each electrode's signal quality, which took around 5--8 minutes depending on the participant.
Participants then completed a six-minute baseline of three two-minute tasks: mental arithmetic, eye movements, and eyes-open rest.
Mental arithmetic raises cognitive load~\cite{harmony1999specific}, eye movements elicit the ocular artifacts that EOG-based correction removes~\cite{croft2000removal}, and eyes-open rest provides a resting reference~\cite{barry2007eeg}.
We used the eyes-open rest as the reference for cognitive load (Section~\ref{sec:measures}).
The participant then worked on the task for 20 minutes while a researcher monitored the EEG recordings.
Participants were not asked to think aloud, as verbalizing their thoughts would have imposed additional cognitive load.
The two tasks were separated by a 10-minute break.
After both tasks, participants were asked to rate their trust in AI using the 5-point Likert TXAI scale~\cite{hoffman2023measures,choudhuri2025guides}.

\subsection{Data}\label{sec:data}

We captured neurophysiological states (RQ1) in 2-second EEG windows, a common approach for spectral EEG measures~\cite{allen2004issues}.
A 2-second window is short enough to fall within a single activity and long enough to estimate theta and alpha power at 1-Hz resolution.
For interactions (RQ2), the units of analysis were transitions between activities, interaction cycles, and prompts (Section~\ref{sec:analysis}).

\textbf{Activity coding.}
To link each EEG window to what participants were doing, two researchers coded the screen recordings continuously with 12 codes (supplemental material~\cite{supplemental}), which make up the four activities we analyzed.
They first coded 20\% of two participants' recordings independently to develop the scheme, and then 20\% of every participant's recordings.
After reaching more than 80\% agreement, they divided the remaining recordings between them.
Each change from one coded segment to the next is a transition (Fig.~\ref{fig:workflow}).
We grouped the codes into four activities in two categories: \emph{articulation} (\WP: composing or revising a prompt) and \emph{verification} (\RAI: reading the AI's response or diffs; \RC: reading code or navigating files; \VA: viewing or interacting with the running application).
We excluded actions related to workflow configuration and idle time.

\textbf{Prompt coding.}\label{sec:prompt-coding}
We transcribed the 141 prompts from the recordings and coded each for its intent type and articulation effort (Table~\ref{tab:prompts}); a prompt could receive more than one code.
From each prompt's intent type, we derived whether it pointed at the code or at the app.
Following Choudhuri et al.~\cite{choudhuri2026copilot}, the codebook includes the code, its definition, and examples, and retains only codes found in prompts from distinct participants.
We coded the prompts with two AI models.
Claude Opus 5.5 derived the codebook and applied it, and GPT-6 independently critiqued and coded all prompts as a second coder alongside Claude.
Two authors then manually reviewed the codebook and spot-checked the coding.
Overall agreement between the two models (Cohen's $\kappa$) was 0.83.

\textbf{EEG processing.}
For each EEG recording, we applied a bandpass filter from 0.5--100\,Hz and a notch filter at 60\,Hz for mains noise.
We screened each recording's power spectrum for flat channels and channels with abnormally high power, and we interpolated the channels we judged unusable on inspection by spherical spline from neighboring sites.
We then removed ocular artifacts using EOG regression and re-referenced the signal to the common average of all channels.
We split each recording into non-overlapping 2-second windows.
Within each window, we estimated each channel's power spectrum with Welch's method (1-second Hann-tapered segments with 50\% overlap, giving 1-Hz resolution) and averaged power over the bins in the theta (4--8\,Hz) and alpha (8--13\,Hz) bands, with 8\,Hz included in both bands.
We rejected windows in which head motion exceeded three standard deviations of the recording's motion signal and windows whose samples were not contiguous in time.
We assigned each window that passed rejection to the coded activity that spanned its full duration, which yielded 7,924 windows from 20 sessions.
We excluded the EEG from two agentic AI sessions on personal projects because the headset's electrodes did not record a usable signal in those sessions.
Their interaction data could still be recovered from the screen recordings and prompts.

\textbf{EEG measures.}\label{sec:measures}
To capture whether participants tended to approach or withdraw from an ongoing activity, which we call \emph{approach--withdrawal} (\aw), we used frontal alpha asymmetry (FAA) ~\cite{harmonjones2018role,coan2004frontal} (Fig.~\ref{fig:workflow}).
Following the right-minus-left convention~\cite{coan2004frontal}, we computed FAA in each window as $\ln\alpha_{F2} - \ln\alpha_{F1}$, where $\alpha_c$ is alpha power at channel $c$.
Because alpha power varies inversely with cortical activity, positive values indicate relatively greater left frontal activity (approach), and negative values indicate withdrawal.
Approach motivation is the impetus to move toward a goal or activity, and withdrawal motivation the impetus to move away from it~\cite{harmonjones2018role}.
A participant invested in their work would therefore show \aw{} that leans toward approach.
Because FAA compares simultaneous recordings from two sites, it does not require a baseline.
To capture \emph{cognitive load}, we used the ratio of frontal theta power to parietal alpha power, which rises with working-memory demand~\cite{gevins2003neurophysiological, holm2009estimating} (Fig.~\ref{fig:workflow}).
In each window $w$, we computed the index $I_w = \bar{\theta}_{F}/\alpha_{Pz}$, where $\bar{\theta}_{F}$ is theta power averaged over F1 and F2, and expressed load relative to eyes-open rest as $\ln(I_w/\tilde{I}_{\mathrm{rest}})$, where $\tilde{I}_{\mathrm{rest}}$ is the median index over the rest windows of the same recording.
Load is therefore 0 at the resting level and positive above it.
\subsection{Analysis}\label{sec:analysis}

\textbf{RQ1: Neurophysiological states.}
To describe participants' neurophysiological states, we report each EEG measure's distribution across windows by AI type, project type, and activity: its median ($M$), its 5th and 95th percentiles ($P_{5}$ and $P_{95}$), which bound the central 90\% of windows, and, for \aw, the share of windows with a strong asymmetry.
We call an \aw{} value strong when it falls outside Tukey's fences~\cite{tukey1977exploratory}, 1.5 interquartile ranges beyond the quartiles.
We computed the fences from all windows pooled across conditions so that they do not depend on AI type.

\textbf{RQ2: Interactions.}
To describe how participants moved between articulating their intent and verifying the AI's work, we computed the transitions between the four analyzed activities, treating consecutive instances of one code as a single \emph{visit}.
We also divided each session into \emph{interaction cycles}.
We write $P$ for \WP and $V=\{O,C,A\}$ for the verification activities \RAI ($O$), \RC ($C$), and \VA ($A$).
A session is a sequence of visits $v_1,\dots,v_n$, each to one activity $a(v_i)\in\{P\}\cup V$; let $t_1<\dots<t_m$ be the positions of the visits to $P$, with $t_{m+1}=n+1$.
Interaction cycle $k$ is the set of verification activities visited between two visits to $P$, $C_k=\{a(v_i) : t_k<i<t_{k+1}\}\subseteq V$.
We kept the 187 cycles that included at least one verification activity ($C_k\neq\emptyset$).
Cycles outnumber the 141 prompts because some visits to \WP ended without a prompt being sent.

To show how close participants came to the implementation when they verified, we ordered the verification activities \emph{from the outside in}: the running app (behavior), the AI's output (its account of what it changed), and the code (the implementation).
For each activity $x\in V$, we computed the share of cycles that included it, $|\{k : x\in C_k\}|$ divided by the number of cycles.
For each cycle, we also recorded the sequence of verification activities in the order participants first reached them after the prompt.
For example, the chain prompt $\rightarrow$ AI output $\rightarrow$ code describes a cycle in which the participant read the AI's output and then the code, without viewing the app (see Fig.~\ref{fig:transitions}).

\textbf{Statistical tests.}
For the interaction measures and the prompt codes, we compared each participant's two sessions by AI type with exact Wilcoxon signed-rank tests~\cite{wilcoxon1945individual}, reporting the magnitude of the matched-pairs rank-biserial correlation ($r$)~\cite{kerby2014simple} as the effect size and reading 0.5 or more as large~\cite{cohen1988statistical}.
Because project type also differed between each participant's two sessions, we assessed its effect by comparing agentic-minus-non-agentic differences across the two groups using exact Wilcoxon rank-sum tests~\cite{mann1947test}, following the analysis of two-period crossover designs~\cite{hills1979crossover}.
To check whether AI type affected the quality of participants' work, we also compared task performance by AI type within each project type, between the two groups, with the same rank-sum test.
Task performance was the proportion of listed features completed on personal projects and the number of automated test cases passed on the legacy codebase (Section~\ref{sec:tasks}).
The tests treat participants as the independent units, so each participant contributed one share per AI type, computed over that session's interaction cycles, transitions, or prompts.
The percentages in Results pool these cycles, transitions, and prompts across participants.
We applied the Benjamini--Hochberg correction~\cite{benjamini1995controlling} within each set of related measures (transitions, the share of interaction cycles that included each activity, and the prompt codes we compared in Table~\ref{tab:prompts}).
Following the American Statistical Association's statement on $p$-values~\cite{wasserstein2016asa}, we treat differences that did not survive correction, and comparisons of the four combinations of AI type and project type, as descriptive.

\subsection{AI usage}

We used Claude Opus 5.5 and GPT-6 to (1) create an initial EEG analysis pipeline, whose code the authors reviewed for correctness; (2) derive and apply the prompt codes (Table~\ref{tab:prompts}) and second-code them for reliability, with the authors reviewing the codebook and spot-checking the coding; and (3) design figures and copyedit.

%% file: figures/workflow.tex
\definecolor{wfNA}{HTML}{1C55A0}     %
\definecolor{wfAG}{HTML}{A9401A}     %
\definecolor{wfInk}{HTML}{1F2A44}    %
\definecolor{wfBase}{HTML}{E3ECF7}   %
\definecolor{wfBaseD}{HTML}{8FA8CF}  %
\definecolor{wfGrid}{HTML}{BCC9E6}   %
\definecolor{wfP}{HTML}{9D8FE6}      %
\definecolor{wfAI}{HTML}{62C2C4}     %
\definecolor{wfC}{HTML}{EDB94F}      %
\definecolor{wfApp}{HTML}{DE6F80}    %
\definecolor{wfCh}{HTML}{4A5878}     %
\definecolor{wfCh2}{HTML}{8A97B5}    %
\definecolor{wfArrow}{HTML}{5B6B8C}  %
\definecolor{wfBox}{HTML}{F4F2FB}    %
\definecolor{wfBoxD}{HTML}{B9B1DE}
\definecolor{wfRQ}{HTML}{E4E8F7}
\definecolor{wfRQD}{HTML}{7D8BC4}
\newcommand{\wfchat}[2]{%
  \begin{scope}[shift={(#1,#2)}]
    \fill[wfNA, rounded corners=0.6mm] (-1.9,-1.2) rectangle (1.9,1.2);
    \fill[wfNA] (-1.0,-1.1) -- (-1.6,-2.0) -- (-0.2,-1.1) -- cycle;
    \foreach \d in {-0.9,0,0.9} \fill[white] (\d,0) circle (0.28);
  \end{scope}}
\newcommand{\wfagent}[2]{%
  \begin{scope}[shift={(#1,#2)}]
    \fill[wfAG, rounded corners=0.6mm] (-1.8,-1.4) rectangle (1.8,1.2);
    \draw[wfAG, line width=0.6pt] (0,1.2) -- (0,1.9); \fill[wfAG] (0,2.0) circle (0.35);
    \fill[white] (-0.75,0) circle (0.38); \fill[white] (0.75,0) circle (0.38);
    \fill[white] (-0.7,-0.85) rectangle (0.7,-0.6);
  \end{scope}}
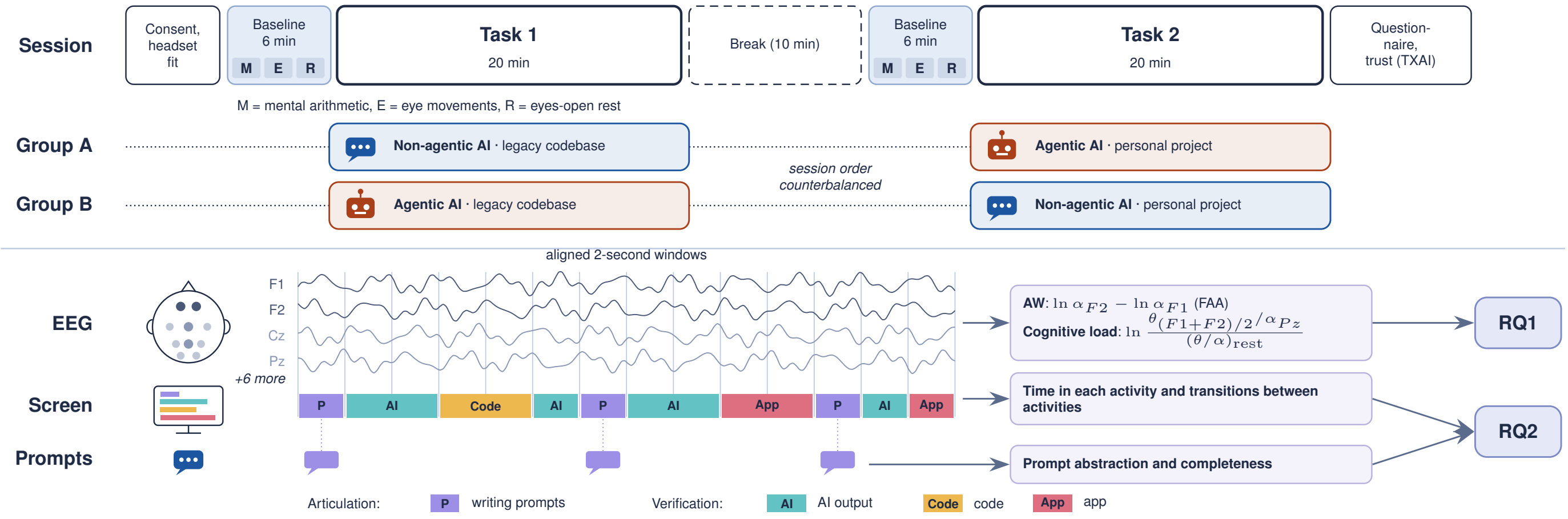
\begin{figure*}[t]
\centering
\resizebox{\textwidth}{!}{%
\begin{tikzpicture}[x=1mm, y=1mm, font=\sffamily\scriptsize, >={Stealth[length=2.6mm, width=1.8mm]},
    rowlab/.style={anchor=east, font=\sffamily\scriptsize\bfseries, text=wfInk},
    chip/.style={rounded corners=1.2mm, line width=0.6pt, minimum height=6mm, minimum width=46mm, inner sep=1.5mm},
    meas/.style={draw=wfInk, rounded corners=1mm, line width=0.4pt, fill=white, align=left, inner sep=1.6mm, anchor=west, font=\sffamily\tiny, text=wfInk},
    rq/.style={circle, fill=wfInk, text=white, font=\sffamily\scriptsize\bfseries, inner sep=1.1mm},
    seg/.style={font=\sffamily\bfseries\fontsize{4.5}{5}\selectfont, text=wfInk}]

\node[rowlab] at (-3,-5) {Session};
\draw[wfInk, rounded corners=1mm, line width=0.5pt, fill=white] (0,0) rectangle (12,-10);
\node[align=center, font=\sffamily\tiny, text=wfInk] at (6,-5) {Consent,\\headset\\fit};
\foreach \xa in {13, 95} {
  \draw[wfBaseD, fill=wfBase, rounded corners=1mm, line width=0.5pt] (\xa,0) rectangle ({\xa+13.2},-10);
  \node[align=center, font=\sffamily\tiny, text=wfInk] at ({\xa+6.6},-3.4) {Baseline\\6 min};
  \foreach \k/\l in {0/M, 1/E, 2/R} {
    \fill[wfBaseD!45, rounded corners=0.3mm] ({\xa+0.6+4.1*\k},-6.6) rectangle ({\xa+4.2+4.1*\k},-9.2);
    \node[font=\sffamily\bfseries\tiny, text=wfInk] at ({\xa+2.4+4.1*\k},-7.9) {\l};
  }
}
\foreach \xa/\name in {27/1, 109/2} {
  \draw[wfInk, rounded corners=1mm, line width=0.9pt, fill=white] (\xa,0) rectangle ({\xa+44},-10);
  \node[font=\sffamily\scriptsize\bfseries, text=wfInk] at ({\xa+22},-3.6) {Task \name};
  \node[font=\sffamily\tiny, text=wfInk] at ({\xa+22},-7.2) {20 min};
}
\draw[wfInk, densely dashed, rounded corners=1mm, line width=0.5pt] (72,0) rectangle (94,-10);
\node[font=\sffamily\tiny, text=wfInk] at (83,-5) {Break (10 min)};
\draw[wfInk, rounded corners=1mm, line width=0.5pt, fill=white] (154,0) rectangle (172,-10);
\node[align=center, font=\sffamily\tiny, text=wfInk] at (163,-5) {Question-\\naire,\\trust (TXAI)};
\node[font=\sffamily\tiny, text=wfInk, anchor=north west] at (13,-10.8) {M = mental arithmetic, E = eye movements, R = eyes-open rest};

\node[rowlab] at (-3,-18) {Group A};
\node[rowlab] at (-3,-25.5) {Group B};
\foreach \y in {-18, -25.5} \draw[wfInk, densely dotted, line width=0.5pt] (0,\y) -- (26,\y) (72,\y) -- (108,\y);
\node[chip, draw=wfNA, fill=wfNA!10] at (49,-18) {};
\wfchat{30}{-18}\node[anchor=west, font=\sffamily\tiny, text=wfInk] at (33,-18) {\textbf{Non-agentic AI} \textperiodcentered\ legacy codebase};
\node[chip, draw=wfAG, fill=wfAG!10] at (131,-18) {};
\wfagent{112}{-18.2}\node[anchor=west, font=\sffamily\tiny, text=wfInk] at (115,-18) {\textbf{Agentic AI} \textperiodcentered\ personal project};
\node[chip, draw=wfAG, fill=wfAG!10] at (49,-25.5) {};
\wfagent{30}{-25.7}\node[anchor=west, font=\sffamily\tiny, text=wfInk] at (33,-25.5) {\textbf{Agentic AI} \textperiodcentered\ legacy codebase};
\node[chip, draw=wfNA, fill=wfNA!10] at (131,-25.5) {};
\wfchat{112}{-25.5}\node[anchor=west, font=\sffamily\tiny, text=wfInk] at (115,-25.5) {\textbf{Non-agentic AI} \textperiodcentered\ personal project};
\node[font=\sffamily\tiny\itshape, text=wfInk, align=center] at (90,-21.75) {session order\\counterbalanced};

\draw[wfGrid, line width=0.5pt] (-16,-31) -- (184,-31);
\foreach \x in {22,28,...,106} \draw[wfGrid, line width=0.35pt] (\x,-34) -- (\x,-52.5);
\node[font=\sffamily\tiny, text=wfInk, anchor=south] at (64,-34) {aligned 2-second windows};

\node[rowlab] at (-3,-40.5) {EEG};
\begin{scope}[shift={(8,-41)}]
  \draw[wfInk, line width=0.5pt, fill=white] (0,0) circle (4.6);
  \draw[wfInk, line width=0.5pt] (-0.9,4.5) -- (0,5.8) -- (0.9,4.5);
  \draw[wfInk, line width=0.5pt] (-4.6,0.9) arc (90:270:0.7 and 0.9);
  \draw[wfInk, line width=0.5pt] (4.6,0.9) arc (90:-90:0.7 and 0.9);
  \foreach \px/\py in {-2.4/0, 2.4/0, -1.6/-2.2, 1.6/-2.2, -1.0/-3.7, 1.0/-3.7} \fill[wfCh2!55] (\px,\py) circle (0.5);
  \fill[wfCh] (-1.0,2.6) circle (0.6); \fill[wfCh] (1.0,2.6) circle (0.6);
  \fill[wfCh2] (0,0) circle (0.6); \fill[wfCh2] (0,-2.2) circle (0.6);
\end{scope}
\foreach \y/\ch/\c/\ph in {-35.5/F1/wfCh/0, -38.8/F2/wfCh/70, -42.1/Cz/wfCh2/140, -45.4/Pz/wfCh2/210} {
  \node[font=\sffamily\tiny, text=\c, anchor=east] at (21.6,\y) {\ch};
  \draw[\c, line width=0.4pt] plot[domain=22:106, samples=200, smooth] (\x, {\y + 0.9*sin(\x*47+\ph) + 0.45*sin(\x*131+2*\ph) + 0.25*sin(\x*151+\ph)});
}
\node[font=\sffamily\tiny\itshape, text=wfInk, anchor=east] at (21.6,-47.6) {+6 more};

\node[rowlab] at (-3,-51) {Screen};
\begin{scope}[shift={(8,-51.2)}]
  \draw[wfInk, rounded corners=0.5mm, line width=0.5pt, fill=white] (-4.4,-2.4) rectangle (4.4,2.6);
  \fill[wfP] (-3.6,1.3) rectangle (-1.2,1.9); \fill[wfAI] (-3.6,0.3) rectangle (2.4,0.9);
  \fill[wfC] (-3.6,-0.7) rectangle (1.0,-0.1); \fill[wfApp] (-3.6,-1.7) rectangle (3.4,-1.1);
  \draw[wfInk, line width=0.5pt] (0,-2.4) -- (0,-3.4) (-1.6,-3.4) -- (1.6,-3.4);
\end{scope}
\foreach \s/\e/\c/\l in {22/28/wfP/P, 28/40/wfAI/AI, 40/52/wfC/Code, 52/58/wfAI/AI, 58/64/wfP/P, 64/76/wfAI/AI, 76/88/wfApp/App, 88/94/wfP/P, 94/100/wfAI/AI, 100/106/wfApp/App} {
  \fill[\c] (\s+0.2,-49.6) rectangle (\e-0.2,-52.6);
  \node[seg] at ({(\s+\e)/2},-51.1) {\l};
}

\node[rowlab] at (-3,-58) {Prompts};
\wfchat{8}{-58}
\foreach \x in {25, 61, 91} {
  \fill[wfP, rounded corners=0.4mm] (\x-2.2,-56.9) rectangle (\x+2.2,-59.1);
  \fill[wfP] (\x-1.2,-59.0) -- (\x-1.8,-60.0) -- (\x-0.3,-59.0) -- cycle;
  \draw[wfP, densely dotted, line width=0.4pt] (\x,-56.9) -- (\x,-52.8);
}

\node[anchor=west, font=\sffamily\tiny, text=wfInk] at (22,-63.6) {Articulation:};
\fill[wfP] (39,-62.5) rectangle (42.6,-64.7); \node[seg] at (40.8,-63.6) {P};
\node[anchor=west, font=\sffamily\tiny, text=wfInk] at (43,-63.6) {writing prompts};
\node[anchor=west, font=\sffamily\tiny, text=wfInk] at (66,-63.6) {Verification:};
\foreach \c/\l/\t/\x in {wfAI/AI/AI output/82, wfC/Code/code/102, wfApp/App/app/116} {
  \fill[\c] (\x,-62.5) rectangle (\x+5,-64.7); \node[seg] at (\x+2.5,-63.6) {\l};
  \node[anchor=west, font=\sffamily\tiny, text=wfInk] at (\x+5.2,-63.6) {\t};
}

\tikzset{mbox/.style={draw=wfBoxD, fill=wfBox, rounded corners=1mm, line width=0.5pt, align=flush left, inner sep=1.6mm, anchor=west, font=\sffamily\tiny, text=wfInk, text width=43mm},
         rqbox/.style={draw=wfRQD, fill=wfRQ, rounded corners=1.5mm, line width=0.6pt, text=wfInk, font=\sffamily\scriptsize\bfseries, minimum width=11mm, minimum height=6.5mm},
         flow/.style={->, draw=wfArrow, line width=0.6pt}}
\node[mbox] (m1) at (113,-40.5) {\textbf{\aw}: $\ln\alpha_{F2}-\ln\alpha_{F1}$ (FAA)\\\textbf{Cognitive load}: $\ln\frac{\theta_{(F1+F2)/2}/\alpha_{Pz}}{(\theta/\alpha)_{\mathrm{rest}}}$};
\node[mbox] (m2) at (113,-50.2) {\textbf{Time in each activity and transitions between activities}};
\node[mbox] (m3) at (113,-58.6) {\textbf{Prompt abstraction and completeness}};
\draw[flow] (107,-40.5) -- (m1.west);
\draw[flow] (107,-50.2) -- (m2.west);
\draw[flow] (95,-58.6) -- (m3.west);
\node[rqbox] (rq1) at (178,-40.5) {RQ1};
\node[rqbox] (rq2) at (178,-54.4) {RQ2};
\draw[flow] (m1.east) -- (rq1.west);
\draw[flow] (m2.east) -- (rq2.west);
\draw[flow] (m3.east) -- (rq2.west);
\end{tikzpicture}}
\caption{Study workflow. Each participant completed two counterbalanced sessions, differing in AI type and project, each after a resting baseline. EEG, screen recordings, and prompts, aligned in time, yield the measures for RQ1 and RQ2.}
\label{fig:workflow}
\end{figure*}

%% file: figures/eeg.tex
\begin{figure}[t]
    \centering
    \includegraphics[width=0.655\columnwidth]{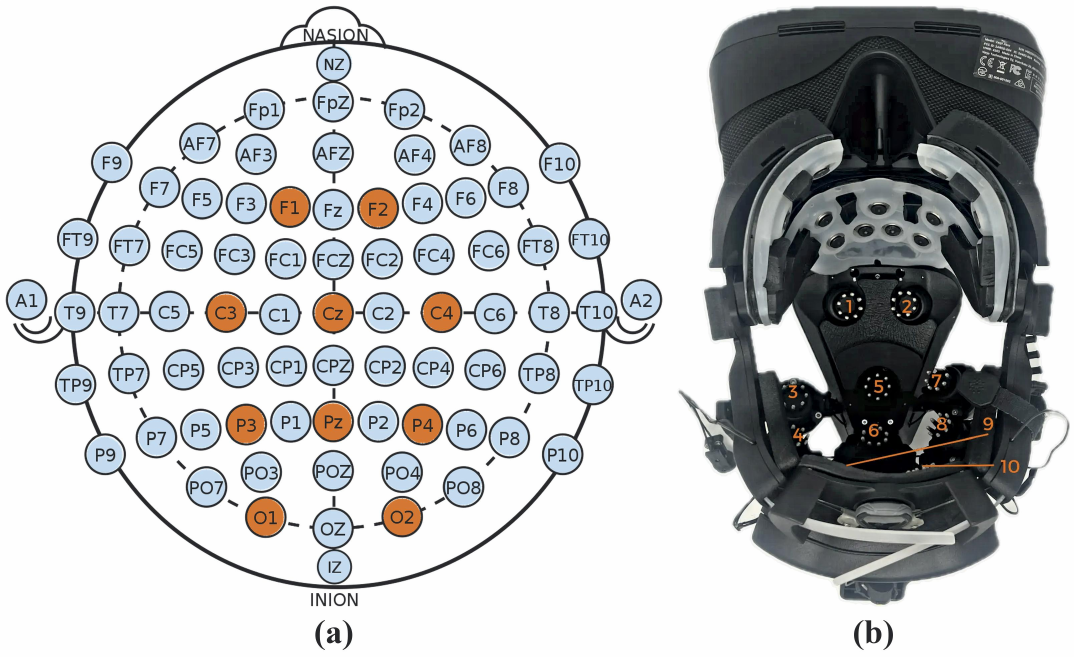}\hfill
    \parbox[b]{0.333\columnwidth}{\centering\includegraphics[width=0.333\columnwidth]{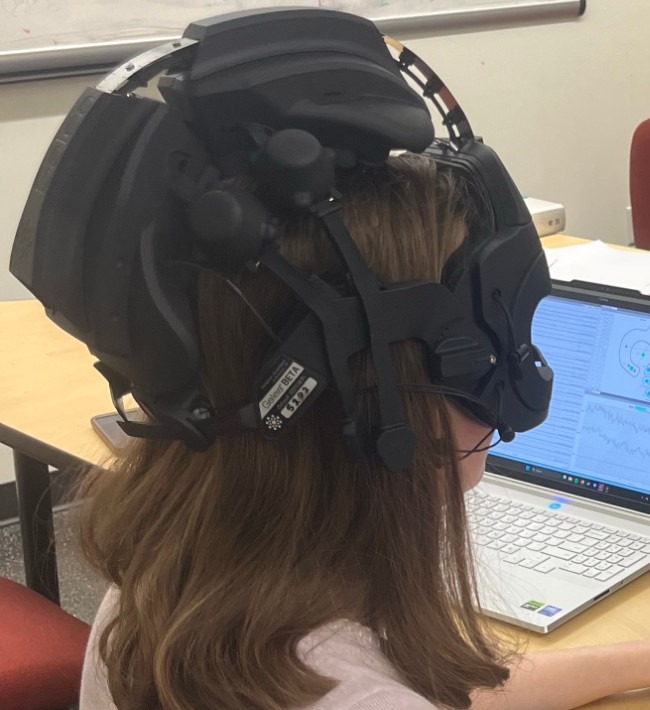}\\[-2.5pt]{\fontsize{6}{7}\selectfont\textbf{(c)}}}
    \caption{(a) The 10--20 system, with the ten sites present in the Galea system highlighted. (b) The OpenBCI Galea Beta~\cite{galea_eeg_sensors}, worn without its VR display, with the electrodes numbered. (c) The headset as worn at a computer.}
    \label{fig:headset}
\end{figure}

%% file: sections/06-threat.tex
\section{Threats to Validity}\label{sec:threats}

\textbf{Construct validity.} We recorded EEG with a wearable dry-electrode headset so that participants could program without gel or lengthy preparation. As Knierim et al.\ note for wearable EEG, such signals are noisier than those from laboratory gel-based systems, and findings from one sensor setup may not transfer to another~\cite{knierim2025exploring}. The headset's frontal pair, F1 and F2, lies next to the F3 and F4 sites that asymmetry studies conventionally use~\cite{coan2004frontal}. We therefore report \aw{} and cognitive load descriptively, over 2-second windows that passed ocular and motion artifact rejection, and interpret distributions across windows. %
Finally, \RC{} does not separate code that participants wrote from code the AI changed, and we did not measure comprehension, so our discussion of program understanding draws on prior work (Section~\ref{sec:discussion}).

\textbf{Internal validity.} Our study had no condition without AI, so we cannot compare participants' states or interactions with programming unaided. We left out such a condition by design: our questions concern vibe coding, in which developers work with AI throughout~\cite{karpathy2025vibe,sarkar2025vibe}, so the comparison of interest is between two ways of working with AI. Each participant's two sessions differed in both AI type and project type, and we separated the two factors in the interaction measures with the standard analysis for crossover designs (Section~\ref{sec:analysis}); because one combination has EEG from only three participants, we describe the EEG measures rather than test them. The two AI types also differed in IDE and default model. Both tools let participants choose any model they offered, as developers can in practice, and participants were asked to let the agent make the changes, as agentic tools are designed to be used. %

\textbf{External validity.} Our $n=11$ participants were students with development experience at one university. A sample skewed toward academia limits generalization to professional developers~\cite{nam2024using}. 20-minute laboratory sessions may not capture how developers work with AI over days~\cite{nam2024using}, brain data vary from day to day~\cite{knierim2025exploring}, and newer agents may change the patterns we observed. We offer these patterns as exploratory evidence of where studies with professional developers, longer periods of use, and newer agents should look.

%% file: sections/04-results.tex
\section{Results}\label{sec:results}

We investigated two beliefs about vibe coding: (1) that developers stay invested when an AI makes the changes, and (2) that delegating to an AI agent lowers their cognitive load.
Our study found that with the agent, participants' \aw{} stayed near neutral and their cognitive load was in fact higher than with non-agentic AI, contradicting both beliefs (Sec.~\ref{sec:rq1}).
To understand these patterns, we then examined how participants interacted with each AI type across the two project types (Sec.~\ref{sec:rq2}).

\subsection{Neurophysiological States Across Treatments (RQ1)}\label{sec:rq1}

\begin{figure*}[t]
    \centering
    \includegraphics[width=\textwidth]{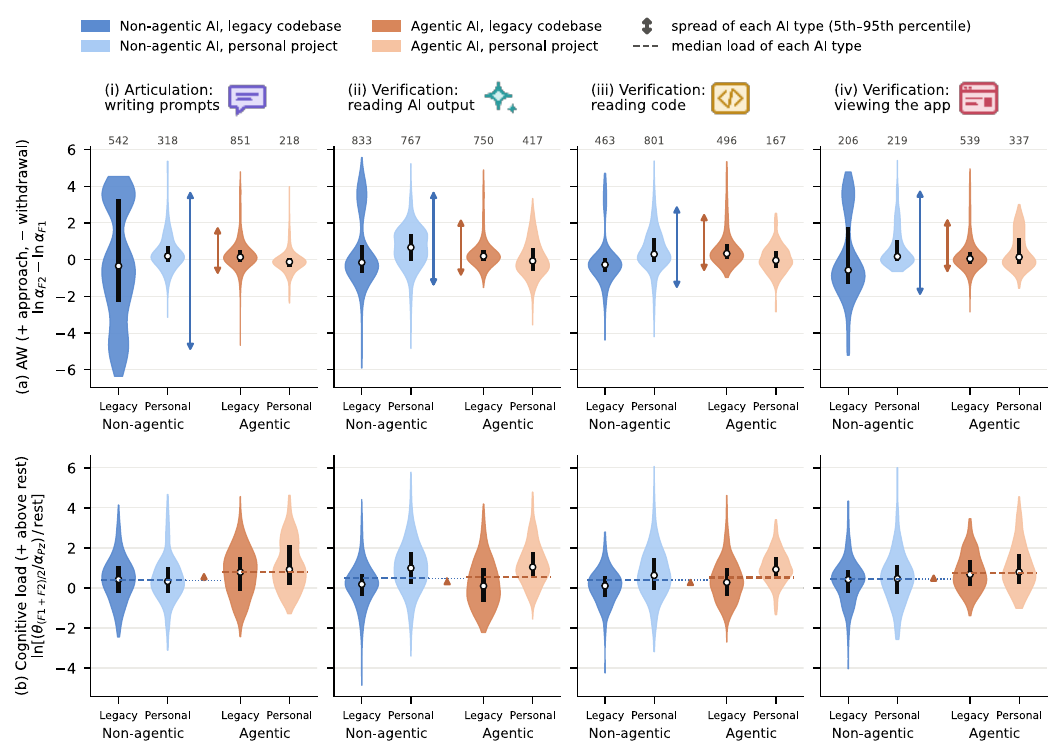}
    \caption{With the agent, \aw{} stayed near neutral and cognitive load was higher than with non-agentic AI. Rows: \aw{} (a) and load (b); columns: activities (i--iv). Each violin shows 2-second EEG windows (counts above row a); white dots mark medians, black bars the interquartile range. Arrows in row a span each AI type's 5th to 95th percentile; dashed lines in row b mark each AI type's median load, and small arrows the step between them. Zero marks symmetry for \aw{} and the resting baseline for load.}
    \label{fig:eeg}
    \vspace{-3mm}
\end{figure*}

Fig.~\ref{fig:eeg} shows participants' \aw{} and cognitive load for each activity (columns i--iv) and each combination of AI type and project type.
If owning the code had kept participants invested, their \aw{} would naturally have leaned toward approach on their own projects regardless of who made the changes.
If delegating had lightened the work, their cognitive load would have been lower with the agent than with non-agentic AI.
However, the data showed otherwise on both counts.

\input{tables/prompt_examples}

\subsubsection{\textbf{Approach--withdrawal}} With the agent, participants' \aw{} stayed near neutral on both project types (own projects: $M=-0.04$; legacy codebase: $M=0.17$) and across every activity (see Fig.~\ref{fig:eeg}a).
Strong asymmetries in either direction were rare with the agent (5\% of windows, compared with 21\% with non-agentic AI), which indicates that participants were largely motivationally indifferent to the outcomes when working with the agent, regardless of project type.
In contrast, with non-agentic AI on their own projects, participants' \aw{} leaned toward approach ($M=0.38$), most of all while they interacted with the AI's output ($M=0.67$; Fig.~\ref{fig:eeg}a(ii)).
Furthermore, with non-agentic AI on the legacy codebase, \aw{} departed strongly toward both approach and withdrawal (35\% of windows showed a strong asymmetry, compared with 7\% on participants' own projects), most of all while participants wrote prompts ($M=-0.34$; Fig.~\ref{fig:eeg}a(i)), suggesting that participants at times leaned toward the work in the legacy codebase and at times withdrew from it.

Understood through the lens of material disengagement~\cite{sarkar2025vibe}, \aw{} departed from neutral when participants worked \emph{with} the AI in a non-agentic mode and stayed near neutral when the AI instead worked \emph{for} them.
In fact, a pharmacological EEG study showed that approach-related frontal asymmetry depends in part on dopamine~\cite{wacker2013dopamine}, and recordings from dopamine neurons in primates showed that these neurons respond to outcomes that are better or worse than expected~\cite{schultz1997neural}, with activity that rises when the outcome is uncertain~\cite{fiorillo2003discrete}.
With non-agentic AI, each change that participants chose and applied had an outcome they waited on, i.e., a step toward their own goal on their own projects, consistent with \aw{} leaning toward approach, and a less predictable step in the legacy codebase, consistent with \aw{} departing toward both approach and withdrawal.
We describe where participants' work and attention went instead in Section~\ref{sec:rq2}, and discuss what the agent's changes did to this reward in Section~\ref{sec:discussion}.

\subsubsection{\textbf{Cognitive load}} With the agent, participants' cognitive load was higher than with non-agentic AI on both project types (own projects: $M=0.94$ vs.\ $0.67$; legacy codebase: $M=0.49$ vs.\ $0.25$; see Fig.~\ref{fig:eeg}b).
Of the four combinations of AI type and project type, load was highest with the agent on participants' own projects.
The difference between the AI types was largest while participants wrote prompts ($M=0.81$ vs.\ $0.39$) and viewed the app ($M=0.74$ vs.\ $0.45$; Fig.~\ref{fig:eeg}b(i) and b(iv)), i.e., while they stated what they wanted and then checked what the app did.
In contrast, while participants read the AI's output and the code, load differed less between the AI types ($M=0.56$ vs.\ $0.49$ and $0.53$ vs.\ $0.40$).
This pattern indicates that with the agent, participants' load was mostly concentrated in stating their intent and checking the app's behavior, which we describe in Section~\ref{sec:rq2}.

Furthermore, load was also higher on participants' own projects than on the legacy codebase with both AI types (non-agentic AI: $M=0.67$ vs.\ $0.25$; agent: $M=0.94$ vs.\ $0.49$), even though participants knew their own code better.
The difference appeared mainly while participants read the AI's output (non-agentic AI: $M=1.00$ vs.\ $0.18$; agent: $M=1.04$ vs.\ $0.10$) and the code (non-agentic AI: $M=0.63$ vs.\ $0.11$; agent: $M=0.93$ vs.\ $0.28$; Fig.~\ref{fig:eeg}b(ii) and b(iii)).
One explanation comes from motivational intensity theory, which holds that involvement of the self makes success important and so justifies mobilizing more effort~\cite{gendolla2010effort}.
By this account, participants may have invested more effort in understanding changes to software that was their own, which would explain this difference.

With the agent on their own projects, participants also faced demands around verifying the agent's changes from the outside in, i.e., running the app and checking its behavior.
In this condition, 86\% of prompts directed changes to the app, compared with 30\% with non-agentic AI on the same project type, whose prompts more often pointed at the code (64\%), and 79\% added a specification, compared with 39\% (see Table~\ref{tab:prompts}).
Such prompts asked the agent for larger changes, and subsequently verifying them in the app came with higher load than with non-agentic AI on the same project type ($M=0.80$ vs.\ $0.47$).
Section~\ref{sec:rq2-prompts} examines these prompts, and Section~\ref{sec:rq2-verify} how participants verified the agent's changes.

\subsection{RQ2: Interactions Across AI and Project Types}\label{sec:rq2}

In RQ2, we investigated how participants' interactions with each AI type differed across the two project types.
Specifically, we report how participants' intent articulation (Section~\ref{sec:rq2-prompts}, Table~\ref{tab:prompts}) and their verification of the AI's work (Section~\ref{sec:rq2-verify}, Fig.~\ref{fig:transitions}) differed across these conditions.
Pennington distinguishes two levels of abstraction at which programmers represent a program: the \emph{program model}, which concerns the code's text and its control flow, and the \emph{domain model}, which concerns what the program does~\cite{pennington1987stimulus}.
Our data showed a similar distinction.
With non-agentic AI, participants' prompts and verification more often concerned the program model, whereas with the agent, they more often concerned the domain model, centering on the app's behavior and outcomes.

\subsubsection{\textbf{Intent articulation moved from the code to the app}}\label{sec:rq2-prompts}

Table~\ref{tab:prompts} lists participants' prompts by what participants asked the AI to do (part A), where their prompts were directed (part B), and what they added alongside the request (part C).

Overall, with the agent, participants' prompts were directed less toward the code and more toward the app: 72\% of prompts on the legacy codebase and 90\% on participants' own projects were directed at the app, compared with 28\% and 33\% with non-agentic AI, differing significantly with a large effect ($p=0.010$, $r=0.89$; see Table~\ref{tab:prompts}B).

Participants also differed in what they asked the AI to do with these prompts.
With the agent, they mostly directed changes to the app's behavior, e.g., ``Make the shadow a little bigger extending to the right more \ldots'' (56\% of prompts on the legacy codebase and 86\% on their own projects, compared with 23\% and 30\% with non-agentic AI).
In contrast, with non-agentic AI, participants more often worked in the code, asking how it worked, e.g., ``what are the paths to these files?'', or asking for code and where to put it, e.g., ``help me write route for cart'' (46\% of prompts across both project types, compared with 6\% with the agent; see Table~\ref{tab:prompts}A).
They also more often intervened in the workflow, approving, advancing, or retrying the AI's work (e.g., ``yes''; ``Next element''), handing over changes to implement, asking for a plan first, and setting the pace and form of its answers (26\%, compared with 12\%).
Because non-agentic AI proposed one change at a time, participants had to tell it when to proceed, whereas the agent carried out multi-step changes on its own.

Finally, participants differed in what they provided alongside their requests.
With non-agentic AI, many prompts were direct requests that stated only what was wanted, e.g., ``help me find new tournament page'' (60\% and 39\%, compared with 25\% and 10\% with the agent; see Table~\ref{tab:prompts}C).
On the other hand, since participants handed the agent whole changes, they added more context to their prompts to the agent, such as what the app was doing then, e.g., ``\ldots it didn't receive and update the value on the screen that I gave'' (48\% of prompts, compared with 26\% with non-agentic AI), alongside more specification, such as constraints on the result, e.g., ``\ldots text should also be larger'' (55\%, compared with 32\%).
Recall from Fig.~\ref{fig:eeg}b(i) that writing prompts was the activity in which load differed most between the AI types ($M=0.81$ vs.\ $0.39$), and this added context and specification coincided with that higher load.

\subsubsection{\textbf{Verification moved from the code to the app}}\label{sec:rq2-verify}

\begin{figure}[t]
    \centering
    \includegraphics[width=\columnwidth]{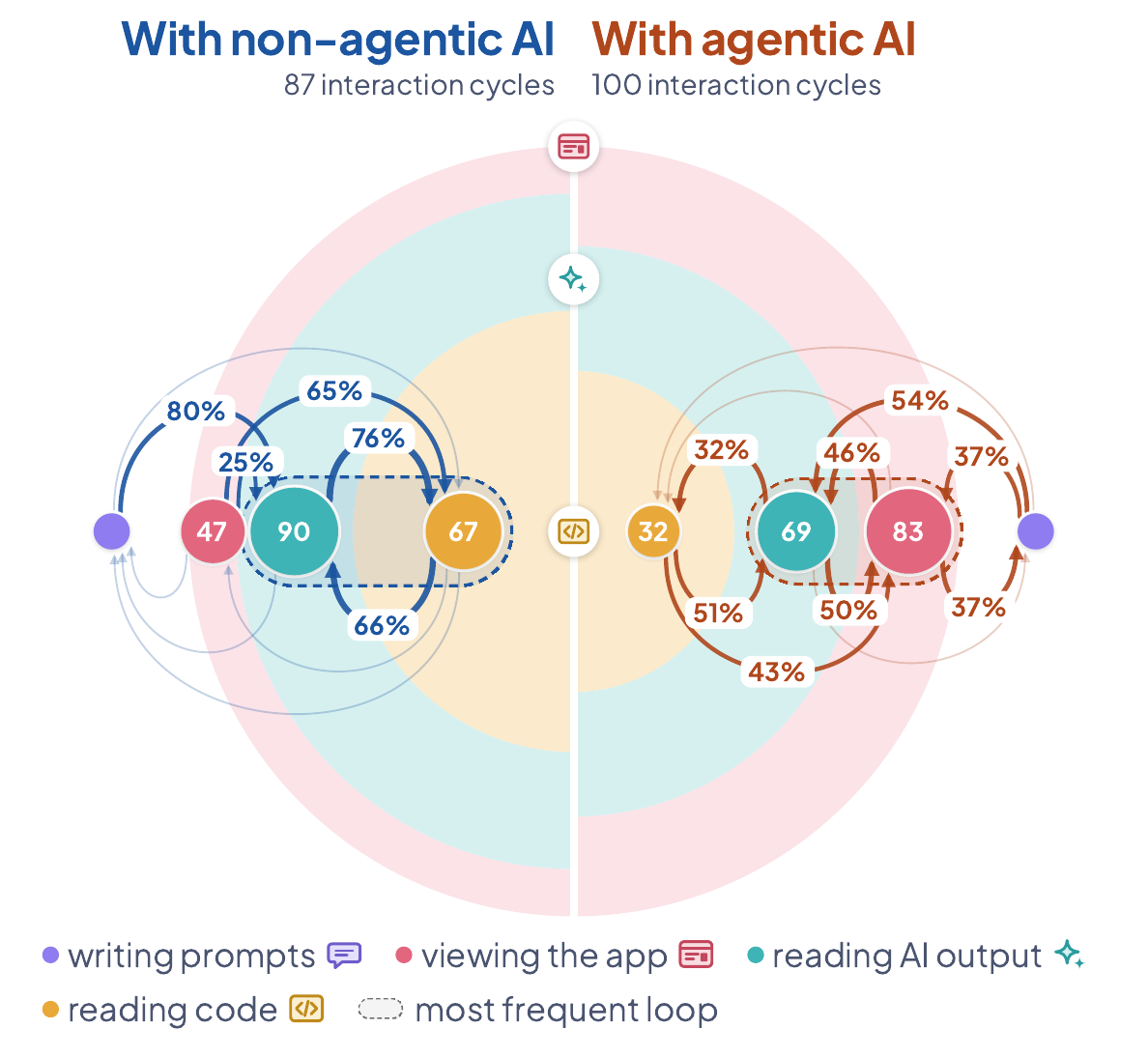}
    \caption{With the agent, participants verified the AI's work in the app more often than with non-agentic AI, and in the code less often. Left: non-agentic AI; right: agentic AI; both project types pooled. Layers run from the app (outside) through the AI's output to the code (center); writing prompts opens each cycle.
    Circle number = \% of interaction cycles that included the activity (area proportional); solid, numbered arrows = $\geq$25\% of the transitions out of an activity; faint arrows = 10--25\%; inward transitions above the axis, outward below; dotted outline = the most frequent loop.}
    \label{fig:transitions}
    \vspace{-3mm}
\end{figure}

Fig.~\ref{fig:transitions} shows how participants verified the AI's work with each AI type.
The three verification activities form layers, from the running app on the outside, through the AI's output, to the code at the center.
Each circle gives the share of interaction cycles that included that activity, each arrow shows how often participants transitioned from one activity to another, and the dotted outline marks the most frequent loop.

Overall, with the agent, participants' verification moved outward, from the code to the app: they read the code in 34\% of interaction cycles on the legacy codebase and 29\% on their own projects, compared with 62\% and 73\% with non-agentic AI, with a statistical significant difference and a large effect ($p=0.006$, $r=0.93$).

Participants also differed in where they went after interacting with the AI's output.
With non-agentic AI, 76\% of these transitions went to the code and 8\% to the app; in contrast, with the agent, 32\% went to the code and 50\% to the app, with the transitions to the code differing significantly with a large effect ($p=0.001$, $r=0.99$).
As a result, the most frequent loop with non-agentic AI ran between the AI's output and the code, whereas with the agent it ran between the AI's output and the app (see the dotted outlines in Fig.~\ref{fig:transitions}).

Furthermore, participants checked the running app more often with the agent, viewing it in 83\% of interaction cycles, compared with 47\% with non-agentic AI.
The most common cycle differed accordingly: with non-agentic AI, participants moved from the prompt to the AI's output and then to the code (26\% of cycles); with the agent, they moved straight from the prompt to the app, bypassing both the AI's output and the code (28\%).
This shift aligned with where participants directed their prompts (recall Section~\ref{sec:rq2-prompts}).

Situation awareness research suggests that people supervising automation must first reconstruct their understanding of the system's state before judging it~\cite{endsley1995outofloop}.
Supervising the agent through the app, a level of abstraction above the code, may have required this reconstruction, contributing to the higher load during app viewing across project types (recall $M=0.74$ vs.\ $0.45$; Fig.~\ref{fig:eeg}b(iv)).
This supervisory role also aligns with participants' motivational indifference: their \aw{} remained near neutral with the agent across project types (Fig.~\ref{fig:eeg}a).

Situation awareness may also help explain the difference in trust between the two groups.
Group A (recall Fig.~\ref{fig:workflow}) used the agent on their own projects, where they were more familiar with the agent's working context, and non-agentic AI on the legacy codebase, where they made atomic changes when working with AI.
Group B had the reverse pairing, coinciding with lower situation awareness.
Consequently, Group B reported higher trust in AI than Group A, although this greater trust was not warranted by task performance, which did not differ significantly between the groups on either project type (see the supplemental material~\cite{supplemental}).
We return to this pattern in Section~\ref{sec:discussion} when discussing over-reliance on AI.

%% file: tables/prompt_examples.tex
\begin{table*}[b]
\centering
\definecolor{naTint}{HTML}{D6E3F4}
\definecolor{agTint}{HTML}{F5DCCF}
\caption{With the agent, prompts were directed at the app more often than with non-agentic AI. Cells give the share of each condition's prompts ($n$) with the code; each intent sums its sub-intents. Shading marks the higher AI type, \colorbox{naTint}{non-agentic} / \colorbox{agTint}{agentic}.}
\label{tab:prompts}
\footnotesize
\renewcommand{\arraystretch}{1.0}
\setlength{\tabcolsep}{3.5pt}
\begin{tabularx}{\textwidth}{@{}>{\raggedright\arraybackslash}p{5.75cm}>{\raggedright\arraybackslash}X rrrr@{}}
\toprule
 & & \multicolumn{2}{c}{\textbf{Non-agentic AI}} & \multicolumn{2}{c@{}}{\textbf{Agentic AI}}\\
\cmidrule(lr){3-4}\cmidrule(l){5-6}
\textbf{A}\enspace Intent type: what the prompt asks for & \textbf{Example} & Legacy (43) & Own (33) & Legacy (36) & Own (29)\\
\midrule
\textbf{Working in the code} &  & \cellcolor{naTint}\textbf{46\%}& \cellcolor{naTint}\textbf{45\%}& \textbf{12\%}& \textbf{0\%}\\
\quad\textit{learning how the code works} & ``what are the paths to these files?'' & \cellcolor{naTint}23\% & \cellcolor{naTint}24\% & 6\% & 0\%\\
\quad\textit{asking for code and where it goes} & ``help me write route for cart'' & \cellcolor{naTint}23\% & \cellcolor{naTint}21\% & 6\% & 0\%\\
\textbf{Directing changes to the app} & ``Make the shadow a little bigger \ldots'' & \textbf{23\%}& \textbf{30\%}& \cellcolor{agTint}\textbf{56\%}& \cellcolor{agTint}\textbf{86\%}\\
\textbf{Reporting what is broken} &  & \textbf{16\%}& \textbf{30\%}& \cellcolor{agTint}\textbf{42\%}& \cellcolor{agTint}\textbf{20\%}\\
\quad\textit{in the app's behavior} & ``the save \& exit button is not working'' & 9\% & 3\% & \cellcolor{agTint}39\% & \cellcolor{agTint}10\%\\
\quad\textit{in errors and test output} & ``fmod is not decleared'' & \cellcolor{naTint}2\% & \cellcolor{naTint}24\% & 0\% & 7\%\\
\quad\textit{in the AI's own work} & ``None of those worked, try reading \ldots'' & \cellcolor{naTint}5\% & \cellcolor{naTint}3\% & 3\% & 3\%\\
\textbf{Intervening in the workflow} &  & \cellcolor{naTint}\textbf{35\%}& \cellcolor{naTint}\textbf{15\%}& \textbf{12\%}& \textbf{13\%}\\
\quad\textit{approving, advancing, or retrying} & ``yes''; ``Next element'' & \cellcolor{naTint}16\% & \cellcolor{naTint}0\% & 3\% & 0\%\\
\quad\textit{handing over changes to implement} & ``Go ahead and implement'' & \cellcolor{naTint}5\% & \cellcolor{naTint}6\% & 6\% & 3\%\\
\quad\textit{asking for a plan first} & ``\ldots can we create a plan to do so'' & \cellcolor{naTint}7\% & \cellcolor{naTint}3\% & 0\% & 3\%\\
\quad\textit{setting the pace and form of the AI's answers} & ``One element at a time. Let's start slow.'' & \cellcolor{naTint}7\% & \cellcolor{naTint}6\% & 3\% & 7\%\\
\midrule
\textbf{B}\enspace Where the prompt points & \textbf{Directed at the code} & \cellcolor{naTint}49\%& \cellcolor{naTint}64\%& 11\%& 7\%\\
 & \textbf{Directed at the app} & 28\%& 33\%& \cellcolor{agTint}72\%& \cellcolor{agTint}90\%\\
\midrule
\multicolumn{6}{@{}l}{\textbf{C}\enspace Articulation effort: what the participant put into the prompt}\\
\textbf{Direct request}: states only what is wanted & ``helpme find new tournament page'' & \cellcolor{naTint}60\%& \cellcolor{naTint}39\%& 25\%& 10\%\\
\textbf{Adds context}: what happens now, what was tried, or why & ``\ldots it didn't receive and update the value on the screen that I gave'' & 21\%& 33\%& \cellcolor{agTint}50\%& \cellcolor{agTint}45\%\\
\textbf{Adds specification}: constraints, criteria, approach, scope, or contingencies & ``\ldots text should also be larger, aim for accessibility heuristics to be followed'' & 26\%& 39\%& \cellcolor{agTint}36\%& \cellcolor{agTint}79\%\\
\bottomrule
\end{tabularx}
\end{table*}

%% file: sections/05-discussion.tex
\section{Discussion and Concluding Remarks}\label{sec:discussion}

\textit{Two Myths of Vibe Coding, Busted!} Much of the case for vibe coding rests on two beliefs: that developers stay invested in their own projects even when they no longer write the code, and that delegating to an agent lowers their load~\cite{sapkota2025vibe,feng2025levels,karpathy2025vibe}. \textit{Our findings contradicted both}.

\subsection{The Vibe Paradox}\label{sec:paradox}

Vibe coding promises that developers can ``forget that the code even exists''~\cite{karpathy2025vibe}. \textit{With the agent, our participants got no relief in cognitive load and showed interest atrophy, and those who used it on the code they knew least reported higher trust in AI.} Together, these three effects are what we term the \emph{vibe paradox}. To explain the paradox, we consider four theories in concert: \emph{supervisory control}~\cite{sheridan1992telerobotics}, \emph{reward from one's own choices}~\cite{yeung2005erp,haggard2017sense}, \emph{motivational intensity}~\cite{gendolla2010effort}, and \emph{automation complacency}~\cite{parasuraman2010complacency}. Fig.~\ref{fig:vibe} shows the paradox as the meeting of these four theories.

\textbf{From working the code to supervising it.} When participants worked with the agent, they no longer made the changes themselves. They told the agent what they wanted the app to do and then checked the result in the running app (Section~\ref{sec:rq2}). Human factors research calls the role this leaves a person in \emph{supervisory control}, instructing an automated system and monitoring it while it carries out the task~\cite{sheridan1992telerobotics}, and Vella argues that coding agents have put software engineers in this role without appropriate work redesign~\cite{vella2026supervisory}.

\textbf{No relief in cognitive load.} Supervisory control explains why participants got no cognitive load relief. Recall that cognitive load was higher with the agent than with non-agentic AI on both project types. Automating a task leaves the human with the harder job of monitoring it and taking over when it fails~\cite{bainbridge1983ironies}, a pattern also found in generative AI~\cite{simkute2024ironies}. With a coding agent, this burden reaches past monitoring the app to stating intent. Indeed, load with the agent exceeded load with non-agentic AI most at exactly these two points (Section~\ref{sec:rq1}).

\input{figures/vibe_paradox}

\textbf{Interest atrophy.} Supervisory control does not explain why participants' \aw{} stayed near neutral with the agent, even in their own code. Approach motivation is tied to reward (Section~\ref{sec:rq1}), and research on \emph{reward from one's own choices} shows that reward-related brain responses are weaker when outcomes do not depend on those choices~\cite{yeung2005erp}, and that the sense of agency fades when actions are not one's own~\cite{haggard2017sense}. With the agent, the choices behind each change were the agent's. We coin \emph{interest atrophy} for this weakening of developers' investment in an outcome when they no longer make the changes themselves. That is, individuals' interest may depend on making the changes as much as on owning the code.

\textbf{Effort without interest.} If reward were the whole story, weaker interest would have meant less effort. Instead, load was highest with the agent on participants' own projects. \emph{Motivational intensity theory} explains this, because involvement of the self makes success important and so justifies effort (Section~\ref{sec:rq1}). Participants still needed the work to succeed, so they paid in effort, though it no longer drew them in.

\textbf{Over-reliance on AI.} In contrast, participants had no such stake in the legacy codebase. Recall that their load there was lower than on their own projects (Section~\ref{sec:rq1}), even though they knew that code least and it may have been the hardest to verify. That is, checking may have cost the most effort where they had the least reason to spend it. Operators who supervise automation come to rely on it without checking its output, a pattern known as \emph{automation complacency}~\cite{parasuraman2010complacency}, and users of AI over-rely on output they cannot verify~\cite{passi2022overreliance}. Consistent with this, participants who used the agent on the legacy codebase reported higher trust in AI than the other participants.

Both beliefs picture an agent that takes the effort off developers' hands and leaves their interest in place. Our participants got the reverse. Delegating took away the part of the work that rewarded them and left them the part that cost them, and where they knew the code least, they reported trusting the AI more.

\subsection{The Vibe Paradox and Solo Software Development}\label{sec:se}

Multi-agent frameworks such as ChatDev and MetaGPT assign the roles of a software team to separate agents~\cite{qian2024chatdev,hong2024metagpt}, making possible what we call solo software development (SSD), in which one developer directs an agent or a team of agents. Ashby's Law of Requisite Variety holds that only variety, the number of different things a system can do, can absorb variety~\cite{ashby1956introduction}. A team of agents can change the process, the product, and what people understand of both in far more ways than one developer can follow, and coordination costs grow faster than the team~\cite{brooks1975mythical}. So a developer stays in control only if they and their tools can absorb the variety the agents produce. Our participants' higher load, and their reading of the code in only a third of cycles, suggest that a prompt and a running app already fall short with one agent. The law gives us three knobs: amplify, attenuate, and maintain.

\textbf{Amplify what the individual can absorb.} Load was higher when participants checked changes in the app, which shows what the program does but not how its parts relate. Tools can show an agent's changes in the structure of the system; for example, Vibe-GUIDE's live graph of a project supported developers' project comprehension relative to a chat-only interface~\cite{chou2026vibeguide}. Tools can also absorb variety themselves by running tests and checks with each change~\cite{vella2026supervisory}, so that only what they cannot settle reaches the developer.

\textbf{Attenuate what the agents produce.} Our two AI types sat at fixed points on the automation dial for a whole session~\cite{parasuraman2000model}. Setting this dial per change lets agents act alone on routine edits but wait on changes that touch a concept central to the system, asking the developer to predict or make them first. Smaller steps leave less to absorb and keep with developers the choices that make an outcome theirs, which developers protect by delegating less of the work tied to their identity~\cite{choudhuri2026shall}.

\textbf{Maintain the match as the work goes on.} Each accepted change moves the system away from what the developer last understood, and a good regulator of a system must be a model of that system~\cite{conant1970good}, so the developer's model has to keep pace with the code. The higher trust among participants who used the agent on the code they knew least suggests that trust in AI can stand in for that model when it is missing. The knob here is how often the agent stops to report. At each stage, agents can state what they intend to change and why, and show which of their changes the developer has inspected in the code, which only in the app, and which not at all. When the report shows that the match has slipped, the developer can turn the other two knobs.

\textit{How far SSD can go depends on how these knobs are set}. We hypothesize that the more agents a developer directs, the more knobs SSD requires: attenuation for a single agent, amplification as well for a small team, and all three for a swarm. Vibe coding promises that developers can forget the code exists. Our findings suggest that this does not hold today, even with one agent. 
To realize this promise and advance the vision of solo software development, where a single human directs a swarm of agents, we need designs grounded in the \textit{law of requisite variety}---that absorb variety at the scale at which agents create it.

%% file: figures/vibe_paradox.tex
\begin{figure}[t]
\centering
\includegraphics[width=0.95\columnwidth]{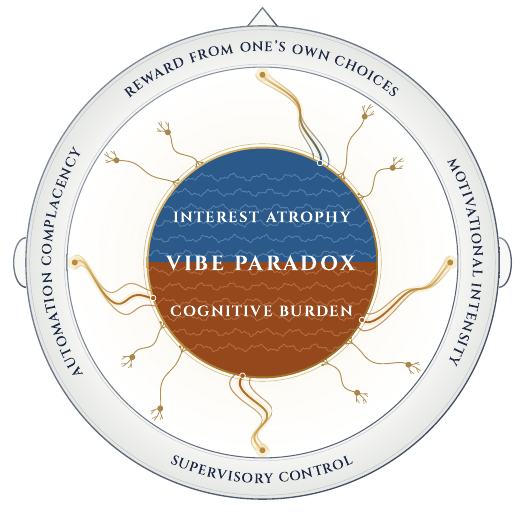}
\vspace{-2mm}
\caption{The vibe paradox. Four theories on the rim shape the two effects at the hub that participants showed with the agent: \emph{interest atrophy} (\aw{} near neutral) and \emph{cognitive burden} (no relief in cognitive load). Reward from one's own choices feeds interest atrophy, and motivational intensity, supervisory control, and automation complacency shape cognitive burden.}
\label{fig:vibe}
\vspace{-2mm}
\end{figure}